\documentclass{elsarticle}
\usepackage{graphicx}% Include figure files
\graphicspath{{figures/pdf/},{figures/eps/}}
\usepackage{dcolumn}% Align table columns on decimal point
\usepackage{bm}% bold math
\usepackage{amsmath,amssymb}

\def\vec#1{\mathbf{#1}}

\def\ket#1{| #1 \rangle}
\def\bra#1{\langle #1 |}

\def\sx{\sigma_x}
\def\sy{\sigma_y}
\def\sz{\sigma_z}

\def\D{\mathcal{D}}

\def\Tr{\mathop{\rm Tr}}
\begin{document}
\DeclareGraphicsExtensions{.pdf}

\begin{frontmatter}
\title{Model discrimination for dephasing two-level systems}
\author[address1,address2]{Er-ling Gong}
\author[address1]{Weiwei Zhou}
%\author[address1]{Zhi-Qiang Sun}
%\author[address1]{Ming Zhang}
\author[address2]{Sophie Schirmer}
\address[address1]{Department of Automatic Control, 
                   College of Mechatronic Engineering and Automation, 
                   National University of Defense Technology, Changsha, Hunan 410073, China}
\address[address2]{College of Science (Physics), Swansea University, 
                   Singleton Park, Swansea, SA2 8PP, United Kingdom}
\date{\today}

\begin{abstract}  
The problem of model discriminability and parameter identifiability
for dephasing two-level systems subject to Hamiltonian control is
studied.  Analytic solutions of the Bloch equations are used to derive
explicit expressions for observables as functions of time for
different models.  This information is used to give criteria for model
discrimination and parameter estimation based on simple experimental
paradigms.
\end{abstract}


\end{frontmatter}

\section{Introduction}

Control of quantum dynamics by means of Hamiltonian engineering is
recognized as a crucial tool for the development of quantum technology
from QIP applications to novel MRI pulse sequences~\cite{QC1,QC2,QC3}.
The effectiveness of most quantum control strategies is conditional on
the existence of accurate models for control design.  The derivation
of such models for systems subject to both control and decoherence is
therefore crucial for the development of effective control strategies,
and techniques for system identification based experimental data play
a important role in finding such models.  This is increasingly being
realized and reflected by a rapidly growing body of literature in the
field of quantum system identification~\cite{Schirmer2004, Cole2005,
  Cole2006, Wang2007, Schirmer2009, Oi2010, CDC2010, Leghtas2012,
  Maruyama2012}.

In this Letter we specifically address the issue of distinguishability
of different plausible models for dephasing two-level systems in the
presence of a nontrivial Hamiltonian via the time evolution of an
observable.  From qubits as building blocks for quantum information
processing~\cite{QIP} to proton spins in magnetic resonance imaging
(MRI) and spectroscopy~\cite{MRI}, dephasing two-level systems are
ubiquitous in many areas of physics and the ability to differentiate
between decoherence models and identify of model parameters based on
simple experimental paradigms for these basic building blocks is an
important task.

\section{Markovian Master Equation and Bloch Equation}

We study a two-level quantum system such as a spin-$\tfrac{1}{2}$
particle or qubit subject to Hamiltonian control and Markovian pure
dephasing.  The state of the system can be described by a density
operator $\rho$, whose evolution is governed by a Lindbladian master
equation
\begin{equation} \label{eq:ME}  %3
 \frac{\partial\rho(t)}{\partial{t}}
 =-\frac{i}{\hbar}[\hat{H},\rho]+\D[V](\rho),
\end{equation}
with the usual Lindbladian dissipation superoperator 
\begin{equation} \label{eq:DV}  %4
  \D[V](\rho) = V \rho V^\dag - \tfrac{1}{2}(V^\dag V \rho + \rho V V^\dag)
\end{equation}
but with unknown Hermitian operators $H$ and $V$.  Broadly, we are
interested in the determination of the operators $H$ and $V$ given
limited or no prior knowledge of the system, with limited control and
measurement resources.  More specifically, we will be interested in
the question of how to discriminate between two types of probable
models and identify the relevant model parameters.

We note here that while Eq.~(\ref{eq:ME}) is a general model to
describe a quantum system subject to Markovian dynamics, we have a
assumed a special form of the dissipation superoperator appropriate
for modelling a two-level system subject to pure dephasing, which can
be described by an Hermitian operator $V$.  With these assumptions we
can, without loss of generality, choose a basis so that either $H$ or
$V$ is diagonal.  We shall choose a basis so that $V$ is diagonal.  As
$V$ is a pure dephasing process and any component proportional to the
identity can be incorporated into the Hamiltonian $H$, we further
assume that $V$ has zero trace.  Thus, $V$ has eigenvalues that occur
in $\pm$ pairs and we can write
\begin{equation} 
 \label{eq:V} 
  V  = \sqrt{\tfrac{\gamma}{2}}\sz, \quad 
 \sz = \begin{pmatrix} 1 & 0 \\ 0 & -1 \end{pmatrix}
\end{equation}
and $\gamma\ge0$.  Under these assumptions the dissipation
super-operator simplifies
\begin{equation}
  \D[\sz](\rho) = \tfrac{\gamma}{2}(\sz\rho\sz-\rho).
\end{equation}

We can further expand the control Hamiltonian with respect to the
Pauli operator basis $\{I,\sx,\sy,\sz\}$ for the $2\times 2$ Hermitian
matrices
\begin{equation}
\label{eq:H}  
  H(t) = \tfrac{\hbar}{2} \left(\alpha I + \omega_z(t) \sz 
         + \omega_x(t)\sx-\omega_y(t)\sy\right),
\end{equation}
where $I$ is the identity operator and
\begin{equation}
  \sx = \begin{pmatrix} 0 &  1\\ 1 & 0 \end{pmatrix}, \quad
  \sy = \begin{pmatrix} 0 & -i\\ i & 0 \end{pmatrix}.
\end{equation}
Terms proportional to the identity give rise only to a global phase
and can be neglected.  Similarly expanding $\rho$ with respect to the
standard Pauli basis
\begin{equation}  \label{eq:rho}  %6-2 
 \rho=\tfrac{1}{2}(I+v_x \sx + v_y \sy + v_z\sz),
\end{equation}
we can recast Eq.~(\ref{eq:ME}) in the common Bloch equation formulation
\begin{equation}
\label{eq:Bloch}  %7
\begin{pmatrix} \dot{v}_x(t)\\ \dot{v}_y(t)\\ \dot{v}_z(t) \end{pmatrix} 
= \begin{pmatrix} -\gamma        & -\omega_z(t) &-\omega_y(t)\\
                   \omega_z(t) & -\gamma        &-\omega_x(t)\\
                   \omega_y(t) & \omega_x(t)  & 0
   \end{pmatrix}
   \begin{pmatrix}
   v_{x}(t)\\ v_{y}(t)\\ v_{z}(t)
   \end{pmatrix},
\end{equation}
where $v_{\alpha}=\Tr(\rho\sigma_{\alpha})$ and we have assumed units
are chosen such that $\hbar=1$.

\section{Model Discrimination and Parameter Estimation Problem}

The general system identification problem for Eq.~(\ref{eq:Bloch}) is
to find all model parameters $\omega_x$, $\omega_y$, $\omega_z$, and
$\gamma$.  This general identification problem may be difficult to
solve, especially when the parameters are time-dependent.  However,
there are interesting special cases.

One such special case is when dephasing acts in the same basis as the
Hamiltonian, i.e., $H$ and $V$ commute, and $\omega_x=\omega_y=0$.
This is the case that is usually assumed without justification.  When
no control is applied and $H$ is simply a static system Hamiltonian
$H_0$ then this is a reasonable assumption.  However, when control
fields are applied the assumption that $H$ and $V$ commute may not be
valid.  Suppose we have a two-level system with
$H_0=\tfrac{1}{2}\omega_0\sz$ that is driven by a constant amplitude
control field giving rise to a control Hamiltonian $H_C = f(t) \sx$ or
$H_C=f(t) \sy$, for example.  Transforming to a rotating frame and
neglecting counter-rotating terms, this gives an effective Hamiltonian
$H^{\rm RWA} = \omega_z \sz + \omega_x \sx$ or $H^{\rm RWA} = \omega_z
\sz + \omega_y \sy$ where $\omega_z=\Delta\omega_0$ is the detuning of
the field from the resonance frequency $\omega_0$ and $\omega_x$ or
$\omega_y$ is the Rabi frequency $\Omega$ of the driving field.  Thus,
assuming that the field does not affect dephasing, the effective
Hamiltonian $H^{\rm RWA}$ and $V$ no longer commute.

From a model identification perspective, an interesting question is
whether the control affects dephasing --- for example, does $V$ act in
the original system Hamiltonian basis, or the new effective
Hamiltonian basis, and to determine the model parameters.  The first
question can be regarded as a model discrimination problem while the
latter is a parameter estimation problem.  Specifically, we are
interested in whether we can discriminate the different cases by
performing a series of simple experiments, and what the best
experimental protocols are.  Motivated by the discussion above, we
specifically consider three different cases:
\begin{itemize}
\item[(1)] $\omega_z \neq 0$, $\omega_x=\omega_y=0$;
\item[(2)] $\omega_x \neq 0$, $\omega_y=\omega_z=0$;
\item[(3)] $\omega_y \neq 0$, $\omega_x=\omega_z=0$,
\end{itemize}
where (a) can be regarded as the case of a two-level system with no
driving fields applied and (b) and (c) as a two-level system
resonantly driven by a constant amplitude field in the $x$-direction
and $y$-direction, respectively.

\section{Experimental Design and Assumptions}

Lack of precise knowledge about the system typically precludes precise
and sophisticated control.  Therefore experimental protocols for
system identification must be kept simple.  In general minimal
requirements for system identification include (1) the ability to
prepare the system in \emph{some} state $\rho_I$ and (2) the ability
to measure some observable $M$ to obtain information about the system.
With regard to assumption (1) we may not know a priori what the state
$\rho_I$ is but it should be possible to repeatedly initialize the
system in the same state by following the same preparation procedure.
In this spirit we make the following assumptions.

\textbf{(1) Initialization.}  %%%%%%%%%%%%%%%%%%%%%%%%%%%%%%%%%%%%%%%
We assume that we are able to prepare the system in some initial
state.  For simplicity we take this to be a pure state $\rho_I =
\ket{\Psi_I(0)}\bra{\Psi_I(0)}$, where $\ket{\Psi_I(0)}$ takes the
form
\begin{equation}
 \label{eq:psi0}
 \ket{\Psi_{I}(0)}
   = \cos\tfrac{\theta_{I}}{2} \ket{0}+\sin\tfrac{\theta_{I}}{2}\ket{1}
\end{equation}
and $\{\ket{0},\ket{1}\}$ denotes an eigenbasis of $V$ --- although this
assumption will be relaxed later.  In practice this preparation might
correspond to letting the system relax to its ground state and
applying a short control pulse.  In the absense of precise knowledge
of the ground state, the resonance frequency of the system and the
coupling strength, the effective rotation angle $\theta_I$ may not be
known initially and we shall see that such a priori knowledge of
$\theta_I$ is not necessary.  We can formally represent the
initialization procedure by the operator $\Pi(\theta_I)$, which is the
projector onto the state $\ket{\psi_I}$.

\textbf{(2) Measurement.} %%%%%%%%%%%%%%%%%%%%%%%%%%%%%%%%%%%%%%%%%%%
We assume the ability to perform a two-outcome projective measurement.
Without loss of generality we can assume the eigenvalues of the
measurement operator to be $\pm 1$ and write
\begin{equation}  \label{eq:M}
  M = M_+ - M_- = \ket{m_+}\bra{m_+} - \ket{m_-}\bra{m_-}.
\end{equation}
We shall assume that the measurement basis states $\ket{m_{\pm}}$ can
be written as
\begin{subequations} %\label{m+-}
\begin{align}
 \ket{m_{+}} &= \cos\tfrac{\theta_{M}}{2}\ket{0}+\sin\tfrac{\theta_{M}}{2}\ket{1}, \\
 \ket{m_{-}} &= \sin\tfrac{\theta_{M}}{2}\ket{0}-\cos\tfrac{\theta_{M}}{2}\ket{1},
\end{align}
\end{subequations}
so that the choice of the measurement can be reduced to suitable
choice of the parameter $\theta_M$, and we shall indicate this by
writing $M(\theta_M)$.

The problem considered here is similar to that considered in
\cite{Cole2006}.  We still assume only a single initial state and
single fixed measurement.  Unlike in \cite{Cole2006}, however, the
initial state and the measurement are not assumed to commute with the
dephasing operator.

%In this context the experimental design problem can be reduced to the
%problem of choosing suitable values $\theta_{I}$ and $\theta_{M}$ to
%identify the system parameters $\gamma$ and $\omega_*$ in each of the
%three cases above.  

\section{Solution of Bloch Equations}

To address the model discrimination and parameter estimation problem we
analytically solve the Bloch equation (\ref{eq:Bloch}) for initial states
of the form (\ref{eq:psi0}) and determine the predicted measurement outcomes
for a measurement of type (\ref{eq:M}) for three different cases.

\subsection{Case 1: $H=\omega\sz$, $V=\sqrt{\tfrac{\gamma}{2}}\sz$}

In this case the Bloch equation (\ref{eq:Bloch}) reduces to
\begin{equation} %\label{5-7b}
\begin{pmatrix}
 \dot{v}_x(t)\\ \dot{v}_y(t)\\ \dot{v}_z(t)
\end{pmatrix} =
\begin{pmatrix}
  -\gamma &-\omega & 0 \\ \omega & -\gamma & 0 \\ 0 & 0 & 0
 \end{pmatrix}
\begin{pmatrix}
  v_x(t)\\ v_y(t)\\ v_z(t)
\end{pmatrix}.
\end{equation}
The solution for the initial state (\ref{eq:psi0}) is
\begin{equation}
\begin{pmatrix}
 v_x(t)\\
 v_y(t)\\
 v_z(t)
\end{pmatrix} =
\begin{pmatrix}
e^{-\gamma{t}}\cos\omega t\sin\theta_I\\
e^{-\gamma{t}}\sin\omega t\sin\theta_I\\
\cos\theta_{I}
\end{pmatrix}
\end{equation}
and applying the binary-outcome projective measurement $M(\theta_M)$
yields the measurement traces $p(t)=\Tr[M\rho(t)]$.
\begin{equation}
\label{eq:meas1}
  p(t) = e^{-\gamma{t}}\cos\omega t\sin\theta_I\sin\theta_M + \cos\theta_I\cos\theta_M.  
\end{equation}

\subsection{Case 2: $H=\omega\sx$, $V=\sqrt{\tfrac{\gamma}{2}}\sz$}

In this case the Bloch equation (\ref{eq:Bloch}) reduces to
\begin{equation}
\label{eq:Bloch1}
\begin{pmatrix}
\dot{v}_x(t)\\
\dot{v}_y(t)\\
\dot{v}_z(t)
\end{pmatrix}
= \begin{pmatrix}
-\gamma    & 0 & 0\\
0 &-\gamma & -\omega \\
0 & \omega & 0
\end{pmatrix}
\begin{pmatrix}
{v}_{x}(t)\\
{v}_{y}(t)\\
{v}_{z}(t)
\end{pmatrix}.
\end{equation}
The solution for the initial state (\ref{eq:psi0}) is
\begin{equation}
\begin{pmatrix}
{v}_{x}(t)\\
{v}_{y}(t)\\
{v}_{z}(t)
\end{pmatrix}
=\begin{pmatrix}
e^{-\gamma t}\sin\theta_{I}\\
\Phi^x_{2}(t)\cos\theta_{I}\\
\Phi^x_{3}(t)\cos\theta_{I}
\end{pmatrix}
\end{equation}
where $\widehat{\omega} =\sqrt{\omega^2-\tfrac{1}{4}\gamma^2}$ and
\begin{subequations} \label{eq:case2}
\begin{align}
\Phi^x_2(t) &= -e^{-\tfrac{\gamma}{2}{t}} \frac{\omega}{\widehat{\omega}}\sin\widehat{\omega} t,\\
\Phi^x_3(t) &=  e^{-\tfrac{\gamma}{2}{t}} \left[\cos\widehat{\omega}t
                                    +\frac{\gamma}{2\widehat{\omega}}\sin\widehat{\omega} t\right].
\end{align}
\end{subequations}
If $\omega^2<\tfrac{1}{4}\gamma^2$ then $\widehat{\omega}$ is purely
imaginary and the sine and cosine terms above are replaced by the
respective hyperbolic functions.  If $\omega^2=\tfrac{1}{4}\gamma^2$,
the expression $\widehat{\omega}^{-1}\sin(\widehat{\omega} t)$ must be
analytically continued.  Applying the binary-outcome projective
measurement $M(\theta_M)$ yields
\begin{equation}
\label{eq:meas2}
   p(t) = e^{-\gamma t}\sin\theta_I\sin\theta_M + \Phi^x_3(t)\cos\theta_I\cos\theta_M.
\end{equation}

\subsection{Case 3: $H=\omega\sy$, $V=\sqrt{\tfrac{\gamma}{2}}\sz$}

In this case the Bloch equation (\ref{eq:Bloch}) reduces to
\begin{equation}
\label{5-11} \begin{pmatrix}
\dot{v}_x(t)\\
\dot{v}_y(t)\\
\dot{v}_z(t)
\end{pmatrix}
=\begin{pmatrix}
-\gamma & 0 &-\omega\\
0 &-\gamma & 0\\
\omega  & 0 & 0
\end{pmatrix}
\begin{pmatrix}
{v}_x(t)\\
{v}_y(t)\\
{v}_z(t)
\end{pmatrix}.
\end{equation}
The solution for the initial state (\ref{eq:psi0}) is
\begin{equation}
\begin{pmatrix}  v_x(t)\\ v_y(t)\\ v_z(t) \end{pmatrix}
 = \begin{pmatrix}
    \Phi_{1}^y(t)\sin\theta_{I} -e^{-\tfrac{\gamma}{2}{t}}
     \tfrac{\omega}{\widehat{\omega}}\sin\widehat{\omega}t \cos\theta_I\\
     0\\
    \Phi_{3}^y(t)\cos\theta_I + e^{-\tfrac{\gamma}{2}{t}} \tfrac{\omega}{\widehat{\omega}}
     \sin\widehat{\omega} t\sin\theta_I
\end{pmatrix}
\end{equation}
where $\widehat{\omega}=\sqrt{\omega^{2}-\tfrac{\gamma^{2}}{4}}$ and
\begin{subequations}
\begin{align}
 \Phi_{1}^y(t) 
  &= e^{-\tfrac{\gamma}{2}{t}} [\cos\widehat{\omega}t-\tfrac{\gamma}{2\widehat{\omega}_{y}}\sin\widehat{\omega}t]\\
 \Phi_{3}^y(t) 
  &= e^{-\tfrac{\gamma}{2}{t}} [\cos\widehat{\omega}t+\tfrac{\gamma}{2\widehat{\omega}_{y}}\sin\widehat{\omega}t]
\end{align}
\end{subequations}

%e^{-\gamma t/2} \left[\cos(\widehat{\omega}t)- \tfrac{\gamma}{2\widehat{\omega}}\sin(\widehat{\omega}t) \right]\sin\theta_I 
%- e^{-\gamma t/2} \tfrac{\omega}{\widehat{\omega}}\sin(\widehat{\omega}t) \cos\theta_I
                                                                                             
Applying the binary-outcome projective measurement $M(\theta_M)$ yields
\begin{equation}
\label{eq:case3}
 p(t) = \alpha_1 e^{-\tfrac{\gamma}{2}{t}} \cos\widehat{\omega}t + \alpha_2 e^{-\tfrac{\gamma}{2}{t}} \sin\widehat{\omega}t
\end{equation}
where the coefficient functions are
\begin{subequations}
\label{eq:case3:coeff}
\begin{align}
 \alpha_1 &= \cos(\theta_I-\theta_M)\\
 \alpha_2 &= \tfrac{\gamma}{2\widehat{\omega}}\cos(\theta_I+\theta_M)+\tfrac{\omega}{\widehat{\omega}} \sin(\theta_I-\theta_M).
 \end{align}
\end{subequations}
As before, if $\omega^2<\tfrac{1}{4}\gamma^2$ then $\widehat{\omega}$
will be purely imaginary and the sine and cosine terms above turn into
their respective hyperbolic sine and cosine equivalents, and if
$\omega^2=\tfrac{1}{4}\gamma^2$, the expression $\widehat{\omega}^{-1}
\sin(\widehat{\omega} t)$ must be analytically continued.

\section{Discussion of Model Discrimination}

\begin{figure*}
\begin{center}
\includegraphics[viewport=0 50 480 350,width=0.49\textwidth]{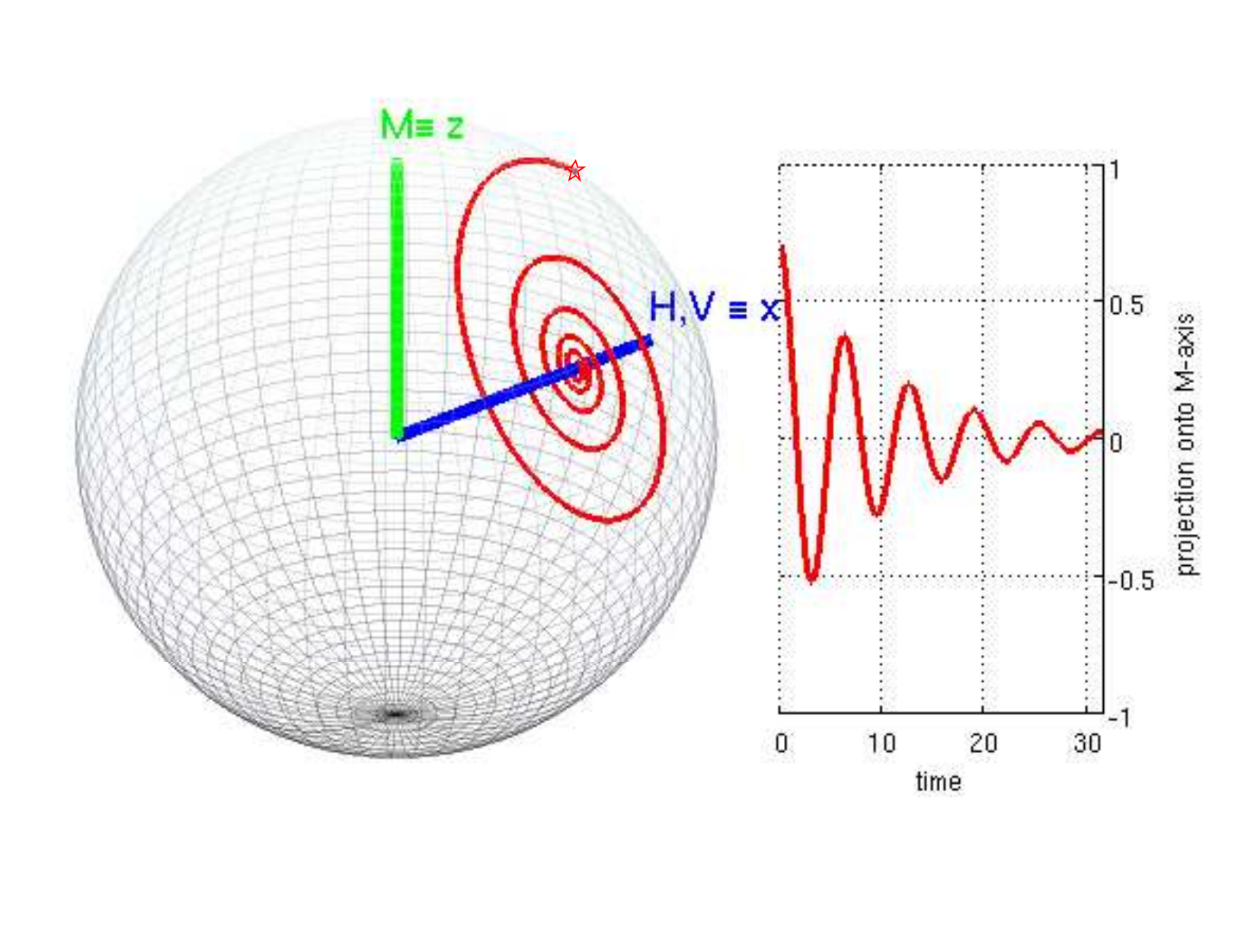}
\includegraphics[viewport=0 50 480 350,width=0.49\textwidth]{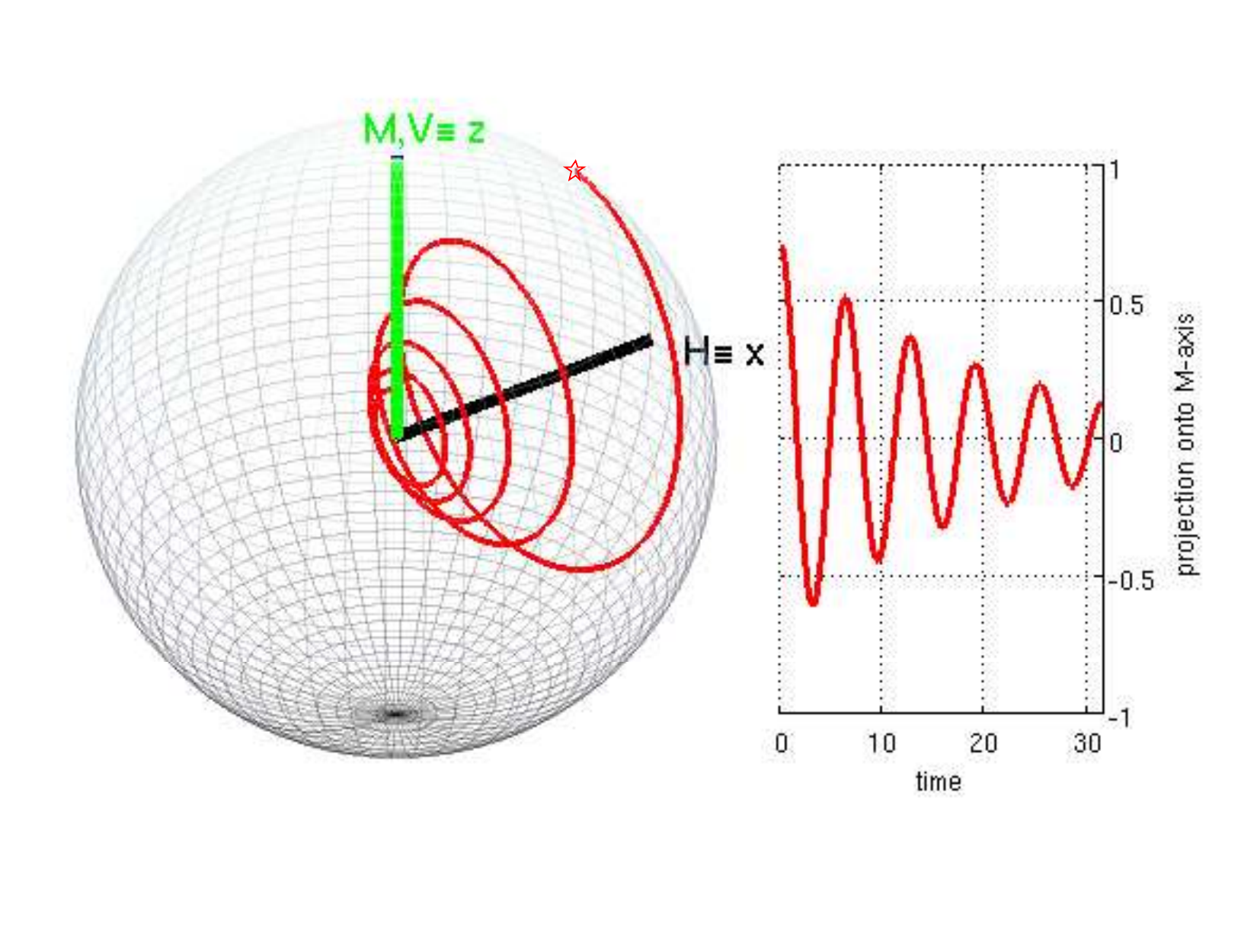}
\end{center}
\caption{Model discrimination problem for $x$-control: Evolution of
  system state on the Bloch sphere and projection onto measurement
  axis for $H\propto\sx$, $\theta_I=\pi/4$, $\theta_M=0$ and
  $V\propto \sx$ (left) and $V\propto \sz$
  (right).}  \label{fig:bloch-x}
\end{figure*}

\begin{figure*}
\begin{center}
\includegraphics[viewport=0 50 480 350,width=0.49\textwidth]{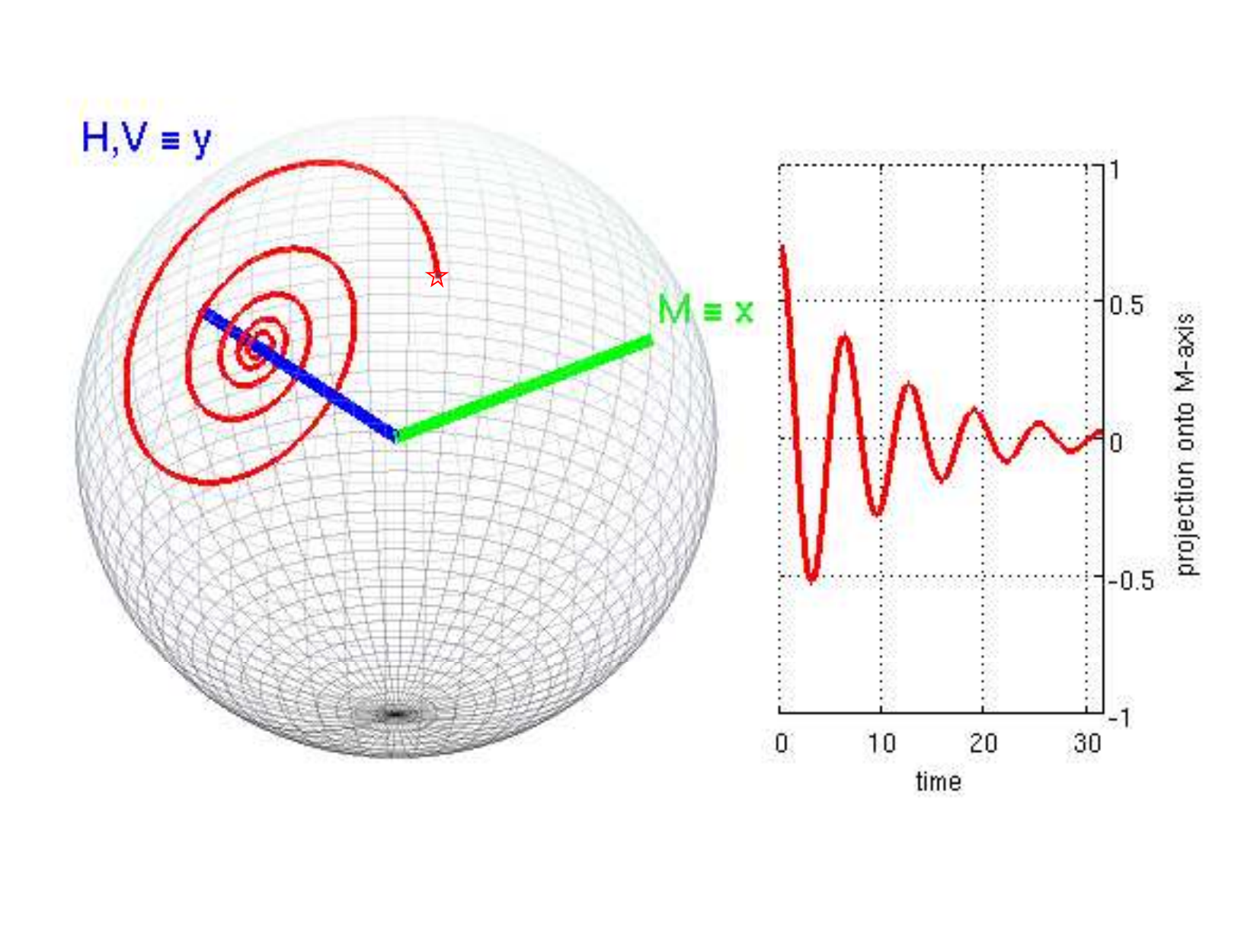}
\includegraphics[viewport=0 50 480 350,width=0.49\textwidth]{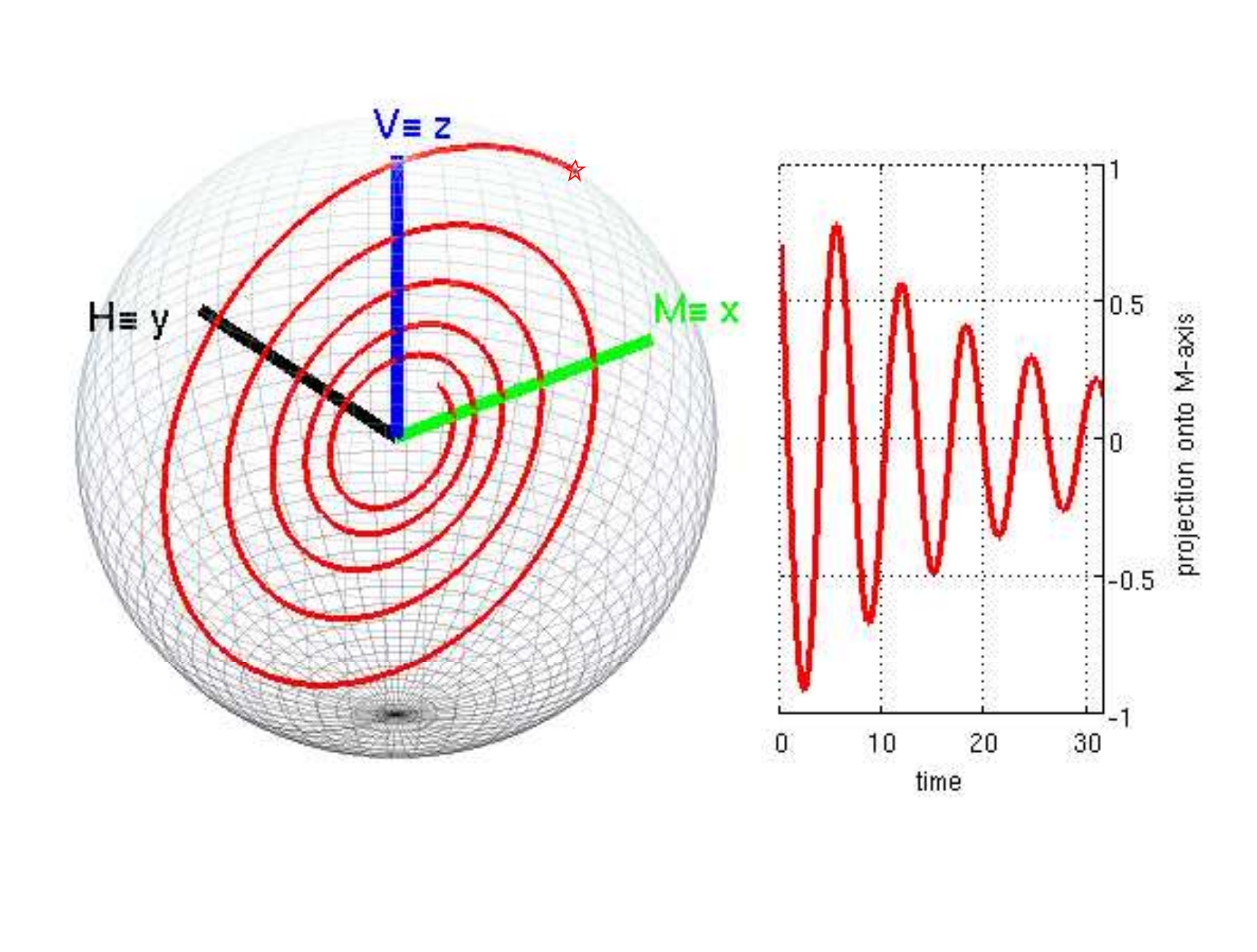}
\end{center}
\caption{Model discrimination problem for $y$-control:
   Evolution of system state on the Bloch sphere and projection
   onto measurement axis for 
   $H\propto\sy$, $\theta_I=\pi/4$, $\theta_M=\pi/2$ and $V\propto \sy$
   (left) and $V\propto \sz$ (right).}  \label{fig:bloch-y}
\end{figure*}

The results of the preceding section show that given the same
initialization and measurement procedures, the different cases lead to
different measurement outcomes:
\begin{subequations}
\begin{align}
p^{(1)}(t) &= e^{-\gamma{t}}\cos\omega t\sin\theta_I\sin\theta_M +\cos\theta_I\cos\theta_M \label{eq:model1a}\\
p^{(2)}(t) &= e^{-\gamma{t}}\sin\theta_I\sin\theta_M +\Phi^{x}_{3}(t)\cos\theta_I\cos\theta_M \label{eq:model2}\\
p^{(3)}(t) &= \alpha_1 e^{-\tfrac{\gamma}{2}{t}} \cos\widehat{\omega}t 
            +\alpha_2 e^{-\tfrac{\gamma}{2}{t}} \sin\widehat{\omega}t \label{eq:model3}.
\end{align}
\end{subequations}
This shows that the models are in principle distinguishable except in
a few special cases.  If $\sin\theta_I \sin\theta_M = \cos\theta_I
\cos\theta_M = 0$ models 1 and 2 are indistinguishable as the
measurement traces for both vanish identically.  This can only happen
if either $\theta_I = m\pi$ and $\theta_M = \tfrac{\pi}{2} + n\pi$ or
vice versa, where $m$ and $n$ are integers.  Models 1 and 3 are always
distinguishable with the given initialization and measurement procedure.

Applied to the problem of distinguishing whether the dephasing acts in
the original basis or the eigenbasis of the new effective Hamiltonian,
a driving field applied in the $x$-direction ($y$-direction) with
dephasing acting in the original ($\sz$) basis corresponds to
\textbf{Case 2} (\textbf{Case 3}) above.  Dephasing acting in the
basis of the new effective system Hamiltonian corresponds to
\textbf{Case 1} above in that both $H$ and $V$ are simultaneously
diagonalizable.  However, we must be careful here as the basis in
which both operators are diagonal \emph{depends on the control field
  applied}, while in the derivation of Case 1 above it was assumed
that both and $H$ and $V$ were diagonal in the $\sz$-basis.  Therefore
a basis change is necessary depending on the direction of the control
field applied.  We explicitly consider the resulting discrimination
problems for three cases.

\textbf{1(a):} If no field is applied or the field is acting in the
$z$-direction then no change of basis is necessary, and the model
discrimination problem is trivial as the effective Hamiltonian acts
in the same basis as the original Hamiltonian.

\textbf{1(b):} If the control is applied in the $x$-direction then the
new effective Hamiltonian is \emph{proportional to $\sx$}, and the
eigenbasis in which $H$ and $V$ are diagonal is $\ket{\pm_x} =
\tfrac{1}{\sqrt{2}}[\ket{0}\pm \ket{1}]$.  We must express the initial
state in this basis
\begin{align}
 \ket{\Psi_I(0)} 
   &= \cos(\tfrac{\theta_{I}}{2}) \ket{0}+\sin(\tfrac{\theta_{I}}{2})\ket{1} \nonumber \\
   &= \cos(\tfrac{\theta_{I}}{2}) \tfrac{\ket{+_x}+\ket{-_x}}{\sqrt{2}}
     +\sin(\tfrac{\theta_{I}}{2}) \tfrac{\ket{+_x}-\ket{-_x}}{\sqrt{2}} \nonumber \\
   &= \tfrac{1}{\sqrt{2}}[\cos(\tfrac{\theta_{I}}{2}) +\sin(\tfrac{\theta_{I}}{2})] \ket{+_x}
     +\tfrac{1}{\sqrt{2}}[\cos(\tfrac{\theta_{I}}{2}) -\sin(\tfrac{\theta_{I}}{2})] \ket{-_x}\nonumber\\
   &= \cos(\tfrac{\theta_I'}{2}) \ket{+_x}
    + \sin(\tfrac{\theta_I'}{2}) \ket{-_x}
\end{align}
with $\theta_I' = \tfrac{\pi}{2}-\theta_I$, and similarly for the
measurement basis states
\begin{equation} 
   \ket{m_\pm} =  \cos(\tfrac{\theta_M'}{2})\ket{+_x} \pm \sin(\tfrac{\theta_M'}{2})\ket{-_x}
\end{equation}
with $\theta_M' = \tfrac{\pi}{2}-\theta_M$, i.e., $\theta_I$ and
$\theta_M$ must be replaced by $\theta_I'$ and $\theta_M'$.  

The \textbf{model discrimination problem for $x$-control}, i.e., the
problem of discriminating whether dephasing acts in the original
Hamiltonian ($\sz$) basis or in the new $\sx$ basis is thus equivalent
to distinguishing
\begin{subequations}
\begin{align}
  p^{(1x)}(t)&= e^{-\gamma{t}}\cos\omega t\cos\theta_I\cos\theta_M +\sin\theta_I\sin\theta_M \label{eq:model1b}\\
  p^{(2)}(t) &= e^{-\gamma{t}}\sin\theta_I\sin\theta_M + \Phi^x_3(t)\cos\theta_I\cos\theta_M.
\end{align}
\end{subequations}
The differences in the evolution and measurement traces between both
cases are illustrated in Fig.~\ref{fig:bloch-x}.  In the first case,
when $H$ and $V$ both act in the $x$-basis, the trajectories follow a
spiral path around the $x$-axis in a plane perpendicular to the
$x$-axis through the point on the sphere defining the initial state,
while in the second case the trajectory follows a spiral path on a
cone around the $x$-axis.  In both cases the measurement signal
corresponds to a damped oscillation but the traces are clearly
distinguishable.

\textbf{1(c)} If a resonant control field is applied in the
$y$-direction then $H^{\rm RWA}\propto\sy$ and the eigenbasis in which
$H$ and $V$ are diagonal is $\ket{\pm_y} =
\tfrac{1}{\sqrt{2}}[\ket{0}\pm i\ket{1}]$.  We must express the
initial state in this basis to be able to apply the results from Case
1 above:
\begin{align}
 \ket{\Psi_I(0)} 
   &= \cos(\tfrac{\theta_{I}}{2}) \ket{0}+\sin(\tfrac{\theta_{I}}{2})\ket{1} \nonumber \\
   &= \cos(\tfrac{\theta_{I}}{2}) \tfrac{\ket{+_y}+\ket{-_y}}{\sqrt{2}}
     -i\sin(\tfrac{\theta_{I}}{2}) \tfrac{\ket{+_y}-\ket{-_y}}{\sqrt{2}} \nonumber \\
   &= \tfrac{1}{\sqrt{2}}[\cos(\tfrac{\theta_{I}}{2})-i\sin(\tfrac{\theta_{I}}{2})] \ket{+_y}
     +\tfrac{1}{\sqrt{2}}[\cos(\tfrac{\theta_{I}}{2})+i\sin(\tfrac{\theta_{I}}{2})] \ket{-_y}\nonumber\\
   &= \tfrac{1}{\sqrt{2}}[\exp(-\tfrac{\theta_I}{2}) \ket{+_y}
    + \exp(+\tfrac{\theta_I}{2}) \ket{-_y}] \label{eq:psi0c}
\end{align}
and similarly the measurement basis states
\begin{equation}
 \ket{m_\pm} =  \exp(-\tfrac{\theta_M}{2}) \ket{+_y} \pm \exp(\tfrac{\theta_M}{2}) \ket{-_y}.
\end{equation}
Solving the Bloch equation (\ref{eq:Bloch1}) for the initial state
(\ref{eq:psi0c}), which has the Bloch vector representation $\vec{v} =
(\cos\theta_I, \sin\theta_I,0)^T$ in the $\sy$ basis, gives
\begin{equation}
\begin{pmatrix}
{v}_{x}(t)\\
{v}_{y}(t)\\
{v}_{z}(t)
\end{pmatrix} =
\begin{pmatrix}
e^{-\gamma{t}}\cos(\omega_zt+\theta_I)\\
e^{-\gamma{t}}\sin(\omega_zt+\theta_I)\\
0
\end{pmatrix}
\end{equation}
and applying the binary-outcome projective measurement $M(\theta_M)$
yields the measurement traces 
\begin{equation}
  p^{(1y)}(t) = \Tr[M\rho(t)]  = e^{-\gamma{t}}\cos(\omega t+\theta_I - \theta_M) .  
\end{equation}

The \textbf{model discrimination problem for $y$-control}, i.e., 
the problem of discriminating whether dephasing acts in the original
Hamiltonian ($\sz$) basis or in the new $\sy$ basis is therefore
equivalent to distinguishing
\begin{subequations}
\begin{align}
  p^{(1y)}(t)&= e^{-\gamma{t}}\cos(\omega t+\theta_I - \theta_M) \label{eq:model1c}\\
  p^{(2)}(t) &= e^{-\gamma{t}}\sin\theta_I\sin\theta_M +\Phi^x_3(t)\cos\theta_I\cos\theta_M.
\end{align}
\end{subequations}
The differences in the evolution and measurement traces between both
cases are illustrated in Fig.~\ref{fig:bloch-y}.  As before, in the
first case, when $H$ and $V$ both act in the $y$-basis, the
trajectories follow a spiral path, this time around the $y$-axis in a
plane perpendicular to the $y$-axis through the point on the sphere
defining the initial state, while in the second case the trajectory
follows a more complex spiral path.  In both cases the measurement
signal corresponds to a damped oscillation but the traces are again
clearly distinguishable.

A driving field applied in an arbitrary direction in the $xy$-plane,
i.e., at an arbitrary angle $\phi_f$ to the $x$-axis, could be
similarly accommodated by a suitable basis change.

\section{Parameter Identifiability}

Once the model type has been identified the next question is if and
how we can identify the parameters in relevant models.

\subsection{Model 1a -- Hamiltonian and dephasing in $\sz$-basis}

\begin{figure*}
\begin{center}
\includegraphics[viewport=0 50 480 350,width=0.49\textwidth]{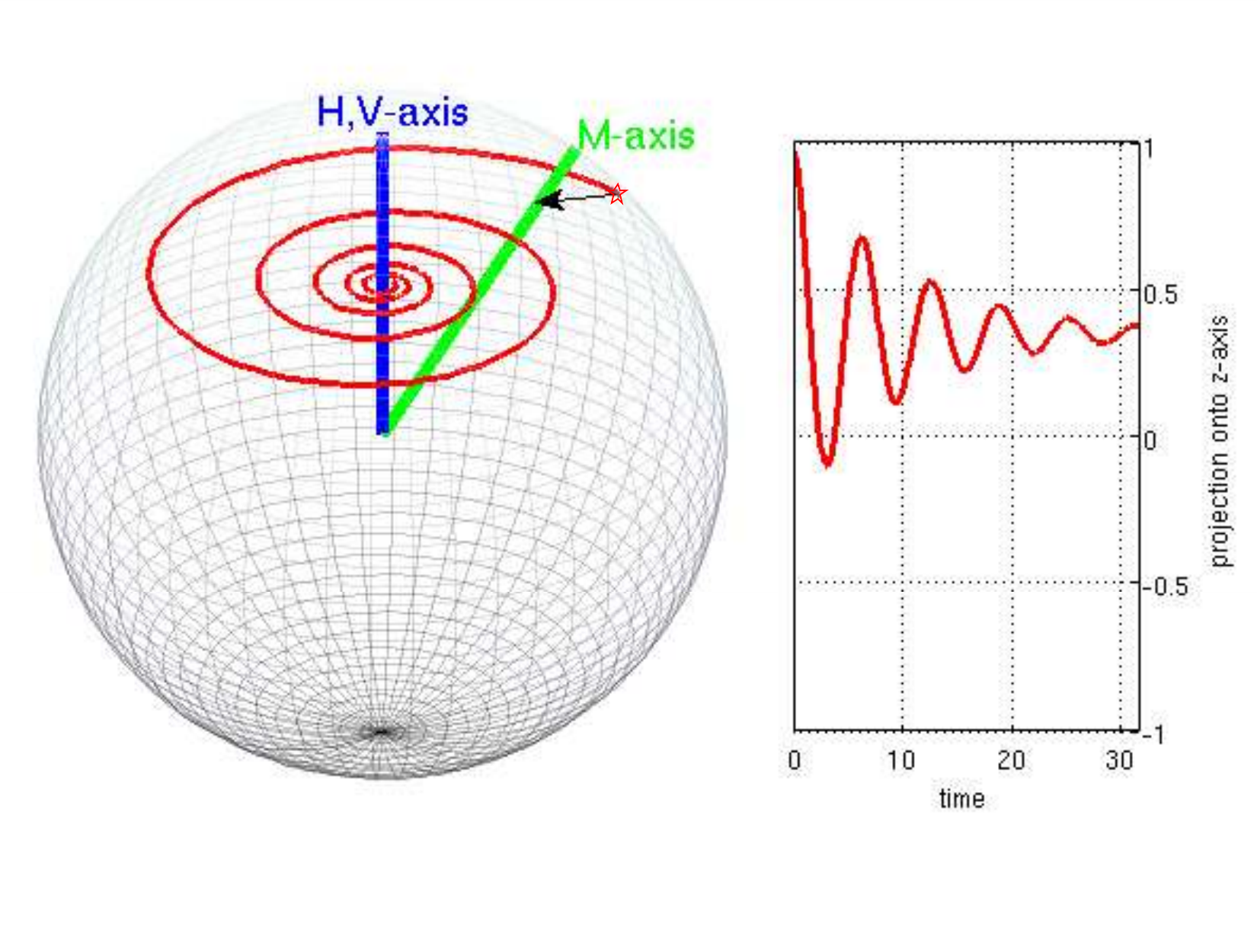}
\includegraphics[viewport=0 50 480 350,width=0.49\textwidth]{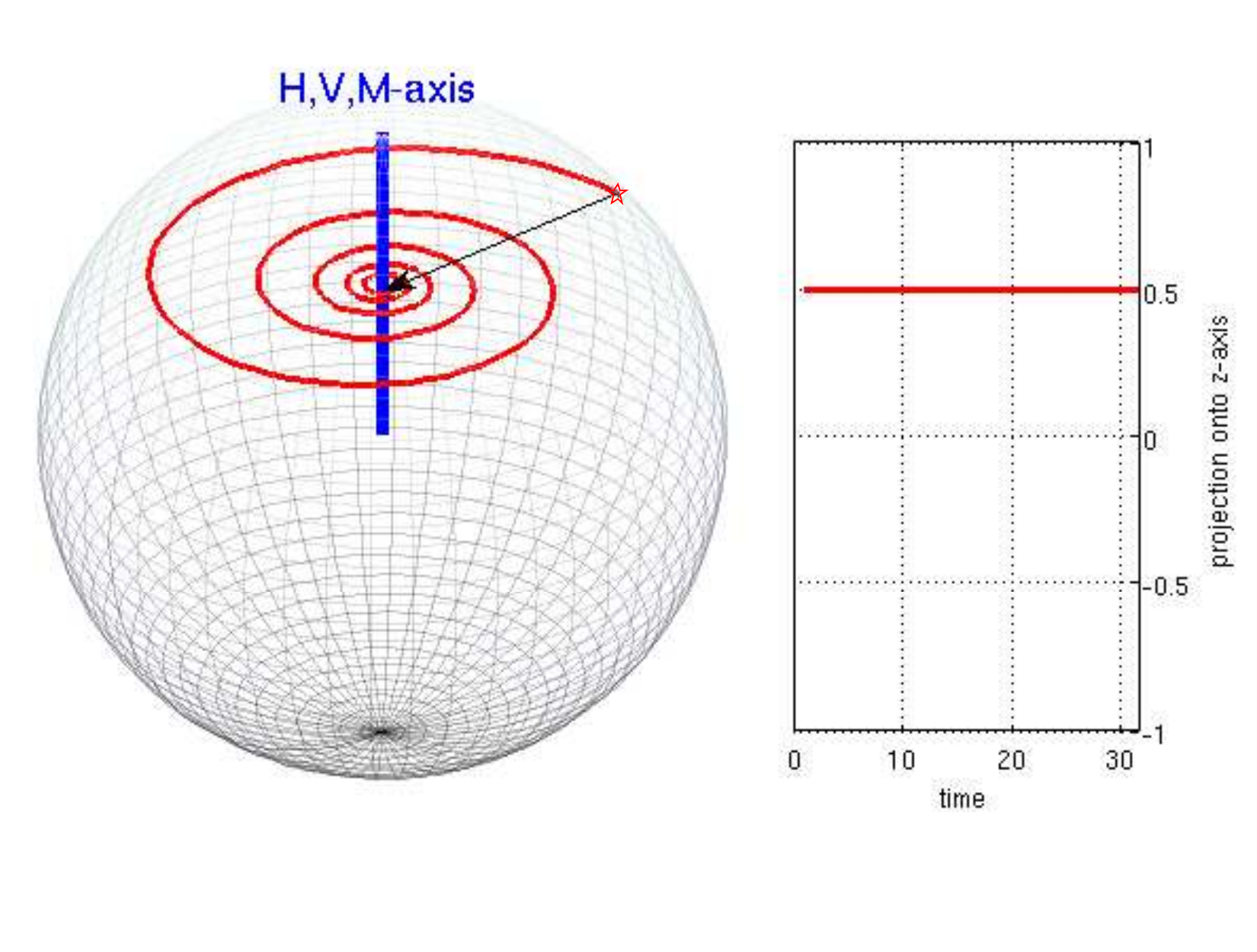}
\end{center}
\caption{Evolution of system state on the Bloch sphere and projection
  onto measurement axis for $H=V$.  The measurement trace contains
  information about both $H$ and $V$ (left), provided the initial
  state is not stationary, and the measurement axis does not coincide
  with the $H,V$-axis.  In the latter case, although the system state
  is not stationary, no information about the system parameters can be
  obtained (right).}  \label{fig:bloch1}
\end{figure*}

Recalling that the observable in case of model (1a) takes the form
\begin{equation}
  p^{(1)}(t) = e^{-\gamma{t}}\cos\omega t\sin\theta_I\sin\theta_M +\cos\theta_I\cos\theta_M \label{eq:model1a}
\end{equation}
shows that we can obtain information about the system parameters
$\omega$ and $\gamma$ if and only if $\sin\theta_I \neq 0$ and
$\sin\theta_M \neq 0$, i.e., if neither the initial state preparation
$\Pi(\theta_I)$ nor the measurement $M$ commutes with $H$ and $V$ as
illustrated in Fig.~\ref{fig:bloch1}.  In this case the measurement
traces also yield information about $\theta_I$ and $\theta_M$, i.e.,
we can determine the relative angles between the initialization and
measurement axis and the fixed Hamiltonian / dephasing axis if they
are not known a priori.  We also see that the visibility is maximized
if $\sin\theta_I\sin\theta_M=1$, which will be the case if the
initialization and measurement axis are orthogonal to the joint
Hamiltonian and dephasing axis.

We can physically understand these results as follows.  As $[H,V]=0$,
if the initial state preparation $\Pi(\theta_I)$ commutes with $H$ and
$V$ then the initial state is a stationary state of the dynamics and
the measurement outcome is constant in time $c_{\pm}=\tfrac{1}{2}(1\pm
1)$.  If $\sin\theta_I\neq 0$ then the initial state is not stationary
and the state follows a spiral path towards the joint Hamiltonian and
dephasing axis but as both $H$ and $V$ are proportional to $\sz$,
$\Tr[\sz \rho(t)]$ is a conserved quantity of the dynamics.  Hence, if
$\sin\theta_M=0$ then the measurement commutes with $\sz$, and as
$\Tr[\sz\rho(t)]$ is a conserved quantity, we are unable to identify
the system parameters $\gamma$ and $\omega_0$, illustrated in
Fig.~\ref{fig:bloch1} (right).  In both cases the constant measurement
result still uniquely identifies the joint eigenbasis of the dephasing
and Hamiltonian operators.

\subsection{Model 1b -- Hamiltonian and dephasing in $\sx$-basis}

If the initial state preparation and measurement operators are
projectors onto states in the $xz$-plane as assumed here then our
derivation above showed that the measurement signal is of the form
\begin{equation*}
  p^{(1x)}(t) = e^{-\gamma{t}}\cos\omega t\cos\theta_I\cos\theta_M +\sin\theta_I\sin\theta_M.
\end{equation*}
This case is similar to the previous case and the change of basis
simply requires adjustment of the initial state and measurement
parameters $\theta_I$ and $\theta_M$.

\subsection{Model 1c -- Hamiltonian and dephasing in $\sy$-basis}

If the initial state preparation and measurement operators are
projectors onto states in the $xz$-plane as we have assumed then the
initial state in this case is always orthogonal to the joint
Hamiltonian and dephasing axis.  This ensures that the model
parameters $\omega$ and $\gamma$ in
\begin{equation*}
  p^{(1y)}(t) = e^{-\gamma{t}}\cos(\omega t+\theta_I - \theta_M) 
\end{equation*}
can be identified for any $\theta_I$ and $\theta_M$ and we always have
maximal visibility.

\begin{figure*}
\begin{center}
\includegraphics[viewport=0 50 561 420,width=0.49\textwidth]{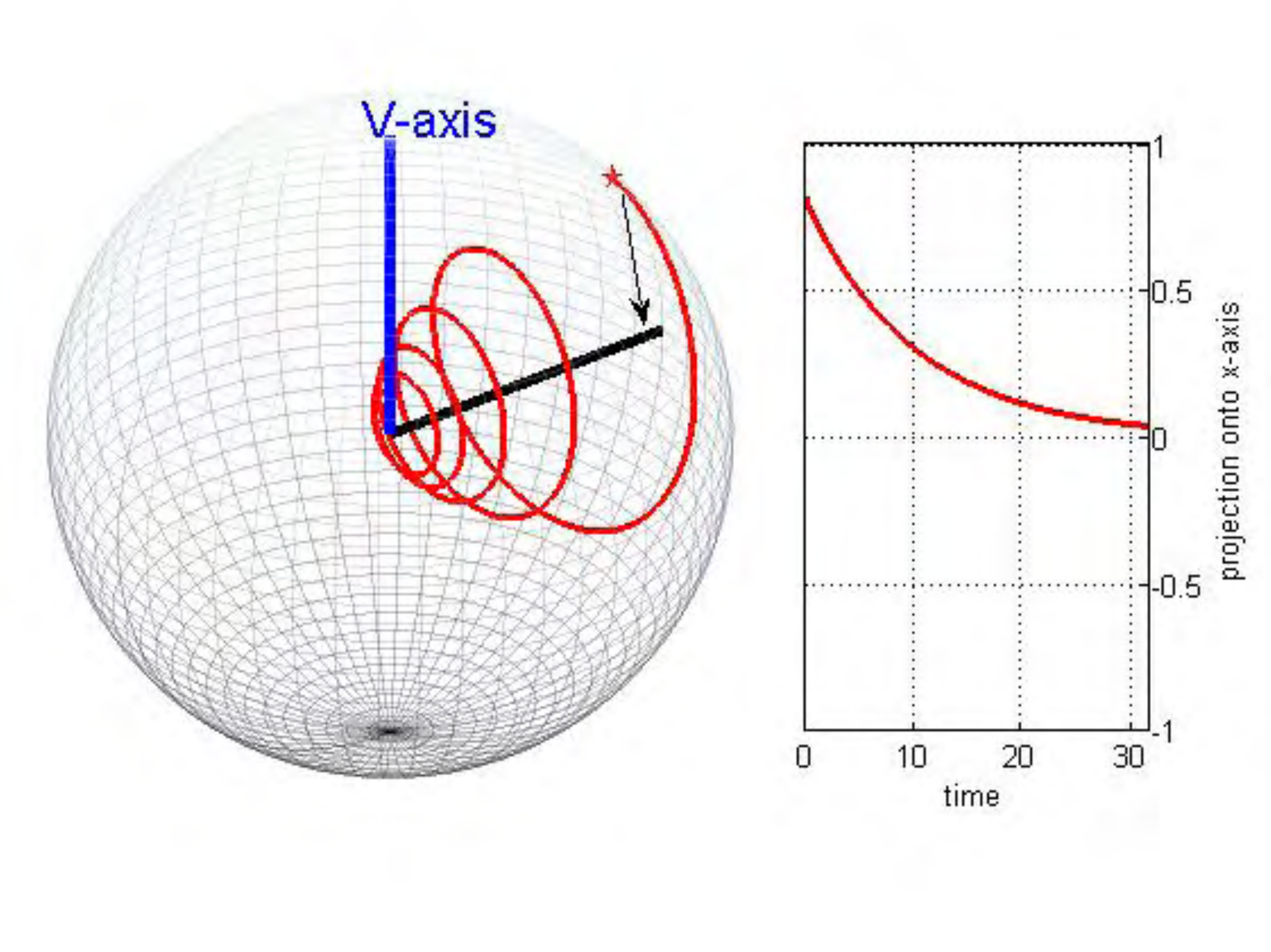}
\includegraphics[viewport=0 50 561 420,width=0.49\textwidth]{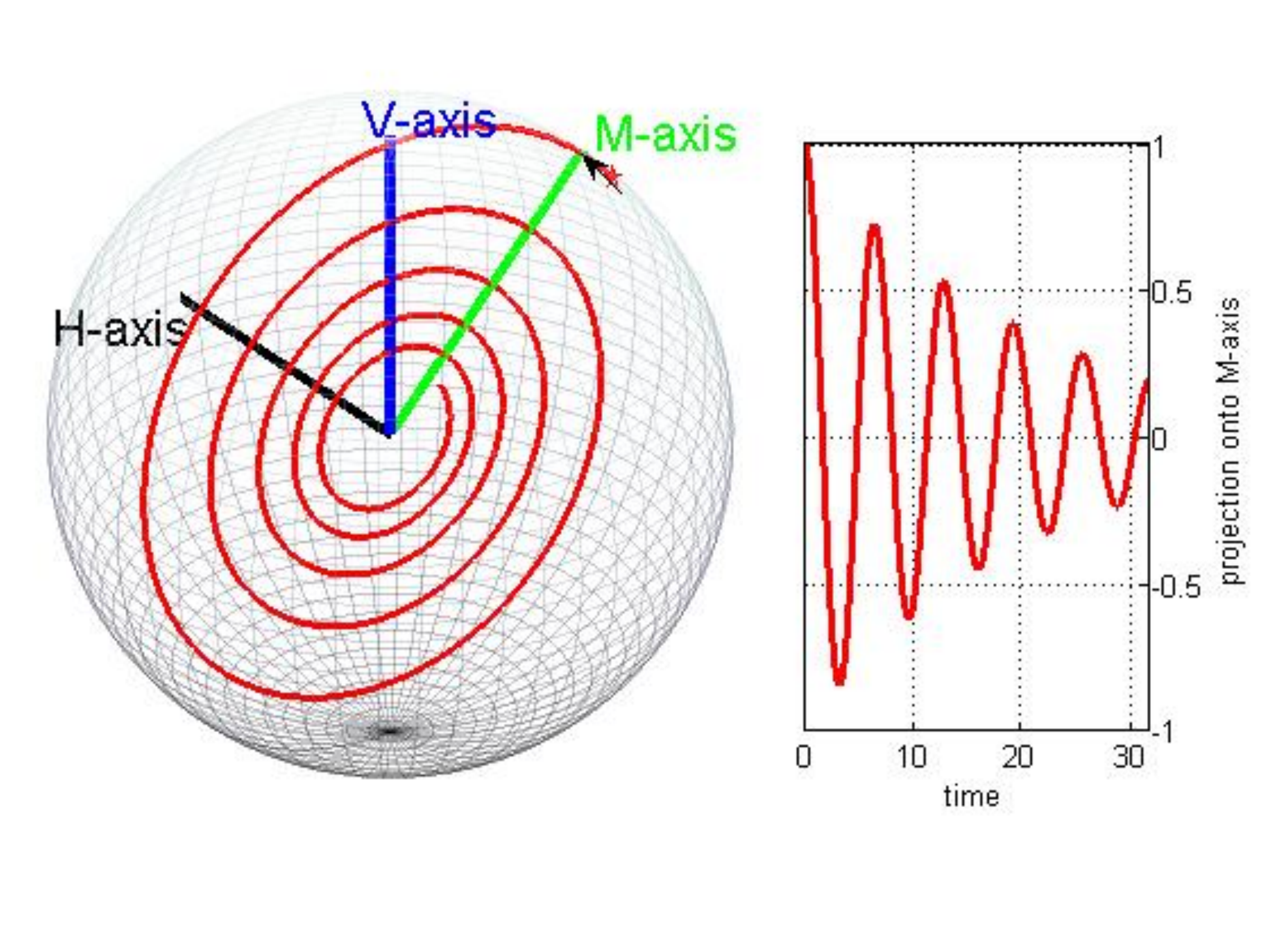}
\end{center}
\caption{Evolution of system state on the Bloch sphere and projection
  onto measurement axis for $H\perp V$: If the measurement axis
  coincides with the $H$-axis, only information about the decoherence
  parameter $\gamma$ can be obtained (left).  If the axes defined by the
  Hamiltonian, dephasing operator and measurement are mutually distinct,
  then the measurement trace always contains information about both $H$ and $V$
  (right).}  \label{fig:bloch2}
\end{figure*}

\subsection{Model 2 -- $\sx$-Hamiltonian, $\sz$-dephasing}

As the expression $\Phi_3^x(t)$ in the measurement signal
\begin{equation*}
  p^{(2)}(t) = e^{-\gamma{t}}\sin\theta_I\sin\theta_M + \Phi^x_3(t)\cos\theta_I\cos\theta_M
\end{equation*}
depends on both $\omega$ and $\gamma$ we can obtain full information
about the system parameters if and only if $\cos\theta_I\neq 0$ and
$\cos\theta_M\neq 0$.  If $\cos\theta_{I}=0$ or $\cos\theta_M=0$, we
can identify $\gamma$ but not $\omega$.  Both cases are illustrated in
Fig.~\ref{fig:bloch2} and can be understood as follows.

$\cos\theta_I=0$ for $\theta_I=\tfrac{\pi}{2}$, i.e., if the initial
state is an eigenstate of the Hamiltonian.  Since $[H,V]\neq 0$ in
this case, eigenstates of $H$ are not stationary.  However, since the
Hamiltonian and dephasing axis are \emph{orthogonal}, the initial
state remains in a plane orthogonal to the dephasing axis, the $z=0$
plane in our case, following the path $x(t)=e^{-\gamma t}$.  Thus we
have $[H,\rho(t)]=0$ for all times, and we can therefore not obtain
any information about the Hamiltonian parameter $\omega$ but we can
still obtain information about the dephasing parameter $\gamma$.  If
the Hamiltonian and dephasing axis were different but not orthogonal
then we would be able to identify both the Hamiltonian and dephasing
parameters even if the initial state was an eigenstate of $H$ as in
this case it would not remain an eigenstate of $H$ under the
evolution.

If $\cos\theta_I \neq 0$ but $\cos\theta_M=0$ then the measurement
commutes with the Hamiltonian.  Transforming to the Heisenberg picture,
\begin{equation*}
  \dot{M}(t) = -i [M(t),H] -\tfrac{\gamma}{2} \D[\sz]M(t),
\end{equation*}
and one can show that $M(t)$ remains orthogonal to the dephasing axis
and $\Tr[M(t)\rho_0]=\Tr[M\rho(t)]$ is independent of the Hamiltonian
$H$, explaining why we cannot obtain any information about $H$ in this
case.

\subsection{Model 3 -- $\sy$-Hamiltonian, $\sz$-dephasing}

It is impossible to find $\theta_{I}$ and $\theta_{M}$ such that
$\cos(\theta_{I}-\theta_{M})
=\sin(\theta_{I}-\theta_{M})=\cos(\theta_{I}+\theta_{M})=0$.  Hence,
we identify both model parameters for any $\theta_I$ and $\theta_M$
from the measurement signal
\begin{equation*}
 p^{(2)}(t) = e^{-\gamma{t}}\sin\theta_I\sin\theta_M+\Phi^x_3(t)\cos\theta_I\cos\theta_M.
\end{equation*}
The reason for this is that the initial state preparation and
measurement operator in this case always project the system into state
in the $xz$-plane which is orthogonal to the Hamiltonian axis,
regardless of the choice of $\theta_I$ and $\theta_M$.  As $[M,H]\neq
0$, there are no conserved quantities and the only stationary state of
the system is the completely mixed state.

\section{Conclusions}

We have studied the problem of model identification for two-level
quantum systems subject to Hamiltonian evolution and simultaneous
dephasing assuming a simple experimental paradigm of repeated
initialization and measurement after a delay.  Analytic solutions of
the Bloch equations were used to derive explicit expressions for the
measured observables for different models.  The latter were used to
elucidate differences between different models and the ability to
discriminate between them experimentally as well as the ability to
identify model parameters from the measurement traces.  The explicit
solutions show that model discrimination and estimation of model
parameters in general do not require a priori knowledge of the
initial state or measurement operators as these can be learned along
with the system parameters from the measurement traces cxcept for a 
few very special (degenerate) cases.

There are practical advantages to formulating the general system
identification problem as a model discrimination problem to decide
whether dephasing acts in the original Hamiltonian basis or the new
effective Hamiltonian basis.  These options appear physically most
plausible, the resulting models are analytically tractable, and
reducing the general model this way, reduces the number of parameters
to be estimated.  Thus, in practice one may wish to start with this
assumption and generalize the problem to allow dephasing in an
arbitary basis if neither model proves to be a good fit to the data.

More work remains to be done to develop efficient algorithms for model
discrimination and parameter identification based on noisy
experimental data.  However, the analytic solutions provide a basis
for the development of optimal experimental protocols and numerical
algorithms.  These tools pave the way for experimental tests of the
validity and accuracy of commonly used models for driven, dephasing
two-level systems, and facilitate the development of more
sophisticated decoherence models to more accurately describe
environmental effects and their interaction with coherent control as
necessary.

\section*{Acknowledgments}
Er-ling Gong acknowledges funding from the China Scholarship Council.
Er-ling Gong and Wei-Wei Zhou also acknowledge funding from the
National University of Defense Technology (China) for research visits
to the UK to enable this joint work and gratefully acknowledge the
hospitality of the University of Cambridge and Swansea University as
host institutions.

%We acknowledge funding from the National Natural Science Foundation of
%China (Grant No 60974037).  SGS acknowledges funding from EPSRC ARF
%Grant EP/D07192X/1 and Hitachi.


\section*{References}

\begin{thebibliography}{apssamp}

\bibitem{QC1} 
H. Rabitz, R. de Vivie-Riedle, M. Motzkus and K. Kompa.
Whither the Future of Controlling Quantum Phenomena?
\textit{Science} \textbf{288} 824 (2000)

\bibitem{QC2}
S. G. Schirmer.  
Quantum choreography: making molecules dance to technology's tune?
\textit{Phil. Trans. R. Soc. A} \textbf{364} 3423--3438 (2006)

\bibitem{QC3} 
D. D'Alessandro, \textit{Introduction to quantum control and dynamics} (CRC Press, 2007)

\bibitem{Schirmer2004} 
S. G. Schirmer, A. Kolli and D. K. L.  Oi. 
Experimental Hamiltonian identification for controlled two-level systems. 
\textit{Phys. Rev. A} \textbf{69} 050306(R) (2004)

\bibitem{Cole2005} 
J. H. Cole, S. G. Schirmer, A. D. Greentree, C. J. Wellard, D. K. L. Oi, and L. C. L. Hollenberg.  
Identifying an experimental two-state Hamiltonian to arbitrary accuracy.  
\textit{Phys. Rev. A} \textbf{71} 062312 (2005)

\bibitem{Cole2006} 
J. H. Cole, A. D. Greentree, D. K. L. Oi, S. G. Schirmer, C. J. Wellard and L. C. L. Hollenberg.
Identifying a two-state Hamiltonian in the presence of decoherence.
\textit{Phys. Rev. A} \textbf{73} 062333 (2006)

\bibitem{Wang2007} 
Z. W. Wang, Y.-S. Zhang, Y.-F. Huang, X.-F. Ren and G.-C. Guo.
Experimental realization of direct characterization of quantum dynamics. 
\textit{Phys. Rev. A} \textbf{75} 044304 (2007)

\bibitem{Schirmer2009} 
S. G. Schirmer and D. K. L. Oi. 
Two-Qubit Hamiltonian Tomography by Baysean Analysis of noisy Data.
\textit{Phys. Rev. A} \textbf{80} 022333 (2009)

\bibitem{Oi2010}
D. K. L. Oi and S. G. Schirmer.
Quantum System Identification by Bayesian Analysis of Noisy Data: Beyond Hamiltonian Tomography. 
\textit{Laser Physics} \textbf{20}(5) 1203-1209 (2010)

\bibitem{CDC2010} 
M. Zhang, S. G. Schirmer, H. Y. Dai, W. Zhou and M. Lin. 
Experimental design and identifiability of model parameters for quantum systems.
\textit{Joint Proceeding of 48th IEEE Conference on Decision and Control and 
        28th Chinese Control Conference}, 3827-3832 (2010)

\bibitem{Leghtas2012}
Z. Leghtas, G. Turinici, H. Rabitz and P. Rouchon.
Hamiltonian Identification Through Enhanced Observability Utilizing Quantum Control.
\textit{IEEE Trans. on Autom. Control} \textbf{57}(10), 2679--2683 (2012)

\bibitem{Maruyama2012}
K. Maruyama, D. Burgarth, A. Ishizaki, K. B. Whaley and T. Takui.
Hamiltonian tomography of dissipative systems under limited access: A biomimetic case study.
Quantum Info. and Comp. Vol. 12, pp. 736-774 (2012)

\bibitem{QIP}
M. A. Nielsen and I. L. Chuang, 
\textit{Quantum Computation and Quantum Information} 
(Cambridge University Press, 2000)

\bibitem{MRI}
Robert W. Brown, E. Mark Haacke, Y.-C. Norman Cheng, Michael R. Thompson and Ramesh Venkatesan,
\textit{Magnetic Resonance Imaging: Physical Properties and Sequence Design}
(John Wiley \& Sons, 2014)

\end{thebibliography}
\end{document}